# Experimental and theoretical investigations on ThGeO$_4$ at high pressure

D. Errandonea[1,†], Ravhi. S. Kumar[2], L. Gracia[3], A. Beltrán[3], S. N. Achary[4], and A. K. Tyagi[4]

[1]MALTA Consolider Team, Departamento de Física Aplicada-ICMUV, Fundación General de la Universitad de Valencia, Edificio de Investigación, c/Dr. Moliner 50, 46100 Burjassot (Valencia), Spain

[2]High Pressure Science and Engineering Center, Department of Physics and Astronomy, University of Nevada Las Vegas, 4505 Maryland Parkway, Las Vegas, Nevada 89154-4002, USA

[3]MALTA Consolider Team, Departament de Química Física I Analítica, Universitat Jaume I, Campus de Riu Sec, Castelló E-12080, Spain

[4]Chemistry Division, Bhabha Atomic Research Centre, Trombay, Mumbai 400085, India

**Abstract:** We report here the combined results of angle-dispersive x-ray diffraction experiments performed on ThGeO$_4$ up to 40 GPa and total-energy density-functional theory calculations. Zircon-type ThGeO$_4$ is found to undergo a pressure-driven phase transition at 11 GPa to the tetragonal scheelite structure. A second phase transition to a monoclinic M-fergusonite type is found beyond 26 GPa. The same transition has been observed in samples that crystallize in the scheelite phase at ambient pressure. No additional phase transition or evidence of decomposition of ThGeO$_4$ has been detected up to 40 GPa. The unit-cell parameters of the monoclinic high-pressure phase are $a = 4.98(2)$ Å, $b = 11.08(4)$ Å, $c = 4.87(2)$ Å, and $\beta = 90.1(1)$, $Z = 4$ at 28.8 GPa. The scheelite-fergusonite transition is reversible and the zircon-scheelite transition non-reversible. From the experiments and the calculations, the room temperature equation of state for the different phases is also obtained. The anisotropic compressibility of the studied crystal is discussed in terms of the differential compressibility of the Th–O and Ge–O bonds.

PACS numbers: 62.50.-p, 61.50.Ks, 61.05.cp, 61.50.Ah, 64.30.-t

I. **INTRODUCTION**

Thorium germanate (ThGeO$_4$) is a member of the ABO$_4$ class of compounds with polymorphism at ambient conditions. ThGeO$_4$ crystallizes either in the tetragonal zircon-type structure (space group: *I4$_1$/amd*) or the tetragonal scheelite-type structure (space group: *I4$_1$/a*) [1]. Both structures are important mineral structures, which consist of AO$_8$ bisdisphenoids and BO$_4$ tetrahedra [2]. The members of the zircon- and scheelite-structured ABO$_4$ family of compounds have gained increasing attention in the past few decades due to their technological applications and their mineralogical interest [3]. In particular, high-pressure (HP) studies have been performed on them, in order to understand their mechanical properties and HP structural behaviour [3 - 10]. Among these studies the majority focus on zircon-type silicates, e.g. ZrSiO$_4$ [8 - 10], and scheelite-type tungstates, e.g. CaWO$_4$ [6, 7]. In both cases, several pressure-induced phase transitions have been discovered and their transition mechanisms were studied [8, 12]. In contrast to these oxides, AGeO$_4$ germanates have been poorly studied upon compression. Among them, only the equation of state (EOS) of scheelite-structured ZrGeO$_4$ and HfGeO$_4$ has been determined up to 20 GPa [13] and a phase transition from zircon to scheelite has been reported in ThGeO$_4$ [14]. Therefore, it is evident that additional research is needed to understand the high-pressure behaviour of ThGeO$_4$ and isomorphic germanates.

In this work, to gain further understanding of the structural properties of orthogermanates, combined HP x-ray diffraction experiments and *ab initio* total-energy calculations on zircon- and scheelite-type ThGeO$_4$ up to 40 GPa are reported. The studies show a zircon-scheelite transition beyond 11 GPa and further a scheelite-monoclinic fergusonite transition (space group: *I2/a*) beyond 26 GPa. Also the EOS and bond compressibility for the different structures is presented. The results are compared with those previously found in other $ABO_4$ compounds. According to the results, $ThGeO_4$ is more compressible than transition metal germanates.

## II. EXPERIMENTAL DETAILS

The experiments were performed on both zircon- and scheelite-structured $ThGeO_4$. The samples used in the experiments were pre-pressed pellets prepared using a finely ground powder obtained from polycrystalline $ThGeO_4$. In order to synthesize scheelite-type $ThGeO_4$, appropriate amounts of pre-heated (1000 ºC) high-purity $ThO_2$ and $GeO_2$ were mixed thoroughly, pelletized, and reheated slowly to 1000 ºC, being held at this temperature for 24 h [1]. Then the pellet was cooled to room temperature (RT), reground, and subsequently heated at 1000 ºC for 15 h. For obtaining zircon-type $ThGeO_4$, the scheelite-type $ThGeO_4$ was heated to 1200 ºC for 24 h, being the scheelite phase transformed irreversibly to zircon-type $ThGeO_4$ [1]. Both products were characterized from their powder x-ray diffraction patterns recorded on a Philips X-Pert Pro diffractometer using monochromatized Cu $K_\alpha$ radiation and by neutron diffraction data collected at the Dhruva Research Reactor at BARC [1]. The refined unit-cell parameters for both phases are given in Table I. They are in good agreement with earlier reported values [1, 15].

Angle-dispersive x-ray diffraction (ADXRD) experiments were carried out on $ThGeO_4$ at RT and HP up to 40 GPa at Sector 16-IDB of the HPCAT - Advanced Photon Source (APS) - using a Mao-Bell type diamond-anvil cell (DAC) with an incident monochromatic wavelength of 0.3447 Å. Samples were loaded in a 100 μm hole of a 40-μm-thick rhenium gasket in the DAC with diamond-culet sizes of 300 μm. Pressure was determined using the ruby fluorescence technique [16] and silicone oil was used as pressure-transmitting medium [17 - 19]. The monochromatic x-ray beam was focused down to 20 × 20 μm$^2$ using Kickpatrick-Baez mirrors. The images were collected using a MAR345 image plate located at 350 mm from the sample. They were integrated and corrected for distortions using FIT2D. The typical exposure time for each spectrum was 20 s. The structural analysis was performed using POWDERCELL.

## III. OVERVIEW OF THE CALCULATIONS

First-principles total-energy calculations were carried out within the periodic density-functional-theory (DFT) framework using the VASP program [20, 21]. The Kohn–Sham equations have been solved by means of the Perdew, Burke, and Ernzerhof (PBE) exchange-correlation functional [21], and the electron–ion interaction described by the projector-augmented-wave (PAW) pseudopotentials [23]. Hybrid density-functional methods have been extensively used for oxides related to $ThGeO_4$, providing an accurate description of crystalline structures, bond lengths, binding energies, and band-gap values [24]. The plane-wave expansion was truncated at a cut-off energy of 400 eV and the Brillouin zones have been sampled through Monkhorst–Pack special $k$-points grids that assure geometrical and energetic convergence for the $ThGeO_4$ structures considered in this work. All the crystal structures are optimized simultaneously on both the volume of the unit-cell and the atomic positions, computing the pressure effect by finding the values of the geometrical parameters that minimize the total energy at a number of fixed volumes. Fittings of the computed energy–volume data with a third-order Birch-Murnaghan EOS [25] provide values of zero-pressure bulk modulus and its pressure derivative as well as enthalpy–pressure curves for the three studied polymorphs [26].

## IV. RESULTS AND DISCUSSIONS

The *in situ* ADXRD data obtained at different pressures, starting from the zircon-type ThGeO4 sample, are shown in Fig. 1. The x-ray patterns could be indexed with the zircon structure up to 8.9 GPa. At 11.1 GPa we found the appearance of diffraction peaks in addition to those assigned to the zircon phase. These peaks can be well assigned to the scheelite structure of $ThGeO_4$, indicating the co-existence of the zircon and scheelite phases from 11.1 GPa up to 12.3 - 13.6 GPa. At 15.8 GPa the diffraction peaks corresponding to the zircon structure disappeared and only the peaks representing the scheelite phase are observed. Upon further compression, the scheelite phase is stable up to 22.3 GPa. The large volume change found at the zircon-scheelite transition (around 10%) indicates that the transition is strongly first order. Additional evidence supporting this conclusion can be found by observing optical changes in small $ThGeO_4$ crystals; microcracks develop beyond 11 GPa altering the transparency of the crystals. Another important fact to be noted here is that pressure induces the zircon-scheelite transition in $ThGeO_4$, but temperature induces a scheelite-zircon transition, as described in the sample preparation details. This fact suggests that there may be an inverse relationship between pressure and temperature in orthogermanates, as previously documented for other $ABO_4$ compounds, like $LaNbO_4$ [27] and $SrMoO_4$ [5].

When compressing the scheelite-structured sample, we found no phase transition from ambient pressure up to 24.4 GPa. Beyond this pressure we found the broadening of diffraction peaks in addition to the appearance of new reflections. The same phenomenon was detected, around 26 GPa, in the scheelite phase of those samples that underwent the zircon-scheelite transition near 11 GPa. These facts indicate the occurrence of a second phase transition in $ThGeO_4$ at 26.2 – 28.8 GPa. The similitude between the diffraction patterns obtained in both cases implies that the post-scheelite phase is the same for those samples that crystallize as scheelite at ambient pressure and those that crystallize as zircon.

The splitting and broadening of the diffraction peaks together with the appearance of new reflections can be seen in Fig. 1 at 28.8 GPa. In particular, the (112) reflection which is the most intense peak of scheelite phase, observed around 2θ = 6.8° at 24.4 GPa, considerably broadens at 28.8 GPa. Similar changes were observed with the (101) reflection located around 2θ = 4.2° at 24.4 GPa. Also the splitting of the (200) reflection of scheelite phase, located near 2θ = 7.9°, becomes visible in Fig. 1. In addition to that, at 28.8 GPa new weak peaks can be clearly observed at 2θ = 3.5° and at 2θ = 11°. These changes seen in the diffraction patterns provide clear evidences for the occurrence of above mentioned structural phase transition. Indeed they resemble those observed at the scheelite-fergusonite transition in other $ABO_4$ compounds [6, 10, 28].

The indexing of the diffraction pattern collected at 28.8 GPa using DICVOL indicates that the post-scheelite structure is monoclinic like that of the M-fergusonite structure (space group: *I2/a*). To evaluate the possibility of assigning it to the post-scheelite structure of $ThGeO_4$, the data was analysed with the LeBail fitting method [29]. Background corrected diffraction patterns could be reasonably well fitted with the LeBail method (with $R_{WP}$ = 3.08 %), supporting the assignment of an M-fergusonite-type structure for the HP phase detected beyond 26.2 GPa. The fitting yielded the unit-cell parameters for the fergusonite structure at 28.8 GPa: $a$ = 4.98(1) Å, $b$ = 11.08(2) Å, $c$ = 4.87 Å, and β = 90.1(1)°. Apparently these parameters imply a small volume change of about 1% during the phase transition. As we will show below, our total-energy calculations also support the scheelite-fergusonite transition. From 28.8 to 40 GPa all the diffraction patterns collected can be assigned to this monoclinic structure, indicating that no additional phase transition takes place in $ThGeO_4$. Furthermore, any possible evidence of the decomposition of $ThGeO_4$ into its component oxides has not been detected. Upon decompression, the scheelite-fergusonite transition is reversible, but the zircon-scheelite transition is irreversible, as can be seen in Fig. 1. This behaviour is typical as documented in $ZrSiO_4$ [9] and $CaWO_4$ [28].

From the refinement of the x-ray diffraction patterns measured up to 24.4 GPa, we extracted the pressure dependence of the lattice parameters and unit-cell volume for zircon and scheelite $ThGeO_4$. These results are summarized in Figs. 2 and 3. As in other zircon-structured $ABO_4$ oxides [11, 30] the compression of zircon $ThGeO_4$ is anisotropic, the *a*-axis being more compressible than the *c*-axis (see Fig. 2). The *c/a* axial ratio increases from 0.904 at ambient pressure to 0.912 at 13.6 GPa. This anisotropy in the axial compressibility of zircon $ThGeO_4$ is comparable with that of zircon-type vanadates [11]. In scheelite $ThGeO_4$ we found the opposite behaviour; the *c*-axis is more compressible than the *a*-axis. In particular the *c/a* axial ratio decreases from 2.241 at ambient pressure to 2.228 at 24.4 GPa. This decrease is typical of scheelite-type oxides [3], however, in the case of $ThGeO_4$ the difference in the axial compressibility is less important than in other scheelites. The difference in the anisotropic behaviour of zircon and scheelite $ThGeO_4$ can be related with the different ordering of $ThO_8$ dodecahedra and $GeO_4$ tetrahedra. The zircon structure can be considered as a chain of alternating edge-sharing $GeO_4$ tetrahedra and $ThO_8$ dodecahedra extending parallel to the *c*-axis, with the chain joined along the *a*-axis by edge-sharing $ThO_8$ dodecahedra [2]. As we will show later, upon compression in both structures the $GeO_4$ tetrahedra are less compressible than the $ThO_8$ bidisphenoids. In the zircon structure, this makes the *c*-axis less compressible than the *a*-axis as observed in our experiments. As a consequence of the symmetry changes between the zircon and the scheelite structure, a rearrangement of the $GeO_4$ and $ThO_8$ units takes place [8]. In particular, in the scheelite structure, the $GeO_4$ tetrahedra are aligned along the *a*-axis, whereas along the *c*-axis the $ThO_8$ dodecahedra are intercalated between the $GeO_4$ tetrahedra. Therefore, in this structure the *a*-axis is the less compressible axis, as found in the experiments.

The pressure dependences of the volume obtained for scheelite and zircon phases are summarized in Fig. 3. There, it can be seen that the transition from zircon to scheelite phase involves a volume collapse of approximately 10%. This is consistent with the volume collapse found in other zircons [8 – 11]. We have analysed the evolution of the volume using a third-order Birch-Murnaghan EOS [26]. The EOS fits for both phases are shown as solid lines in Fig. 3. The obtained EOS parameters for the zircon phase are: $V_0$ = 341.8(9) Å$^3$, $B_0$ = 184(6) GPa, and $B_0$' = 4.6(5), these parameters being the zero-pressure volume, bulk modulus, and its pressure derivative, respectively. The bulk modulus of zircon-type ThGeO$_4$ is 15% smaller than that of ZrSiO$_4$ [9, 31], but larger than that of zircon-type vanadates [10]. The EOS parameters for the scheelite phase are: $V_0$ = 305.2(8) Å$^3$, $B_0$ = 186(6) GPa, and $B_0$' = 4.7(5). This indicates that scheelite-type ThGeO$_4$ is more compressible than scheelite-type ZrGeO$_4$ and HfGeO$_4$ [13]. Similar differences are observed when comparing the compressibility of transition metal vanadates (e.g. ScVO$_4$) with lanthanide vanadates (e.g. EuVO$_4$) [11]. This could be probably related to the fact that transition metals make stronger bonds with oxygen than *f*-electron elements like actinides and lanthanides [32].

Empirical models have been developed for predicting the bulk moduli of zircon-structured and scheelite-structured ABO$_4$ compounds [28]. In particular, the bulk modulus of ThGeO$_4$ can be estimated from the charge density of the ThO$_8$ polyhedra using the relation $B_0 = 610\, Z_i / d^3$, where $Z_i$ is the cationic formal charge of thorium, $d$ is the mean Th-O distance at ambient pressure (in Å), and $B_0$ is given in GPa [28]. Applying this relation a bulk modulus of 170(25) GPa is estimated for zircon ThGeO$_4$ and a bulk modulus of 175(26) GPa is estimated for the scheelite-type phase. These estimations reasonably agree with the values obtained from the experiments and indicate that the scheelite-type phase is slightly less compressible than the zircon-type phase. This is in agreement with the fact that scheelite provides a more efficient atomic packing than zircon.

Let's compare now the experimental data presented above with the results of the *ab initio* calculations for ThGeO$_4$. The zircon, scheelite, and M-fergusonite structures have been considered in these calculations to test the experimental results. Figure 4 shows the energy vs volume curves for these structures. The common tangent construction enables to deduce the transition pressure and the equilibrium pressure [33, 34]. According to the calculations zircon is the most stable structure from ambient pressure up to 2 GPa. Beyond this pressure the scheelite structure become energetically more favorable, which agrees with the zircon-scheelite phase transition detected in the ADXRD experiments. The transition-pressure difference between experiments and calculations may be possible due to a kinetic hindrance of the equilibrium phase transformation, a frequent phenomenon in ABO$_4$ oxides [6], which in some cases leads to a polymorphism zone in the P-T phase diagram [35].

For the zircon structure at ambient pressure, the calculations gave $a$ = 7.3269 Å and $c$ = 6.6416 Å. The obtained atomic positions are summarized in Table I. The calculated unit-cell parameters are slightly larger than the experimental values (see Table I). This small overestimation is within the typical reported systematic errors in DFT calculations. Regarding the atomic positions, the agreement between theory and experiment is very good. The calculated EOS of zircon ThGeO$_4$ is given by the following parameters $V_0$ = 365.54 Å$^3$, $B_0$ = 158.95 GPa, and $B_0$' = 4.14. The value of the bulk modulus slightly underestimates the experimental value (184 GPa) and that obtained from empirical estimations (170 GPa) done following Ref. 28. However, the differences are within the typical reported systematic errors in DFT calculations [34].

On the other hand, our calculations give an anisotropic compressibility for the unit-cell parameters comparable with the experiments.

As pressure increases, the zircon structure becomes unstable against scheelite at 2 GPa. For the scheelite structure at ambient pressure, the calculations gave $a = 5.2128$ Å and $c = 11.6022$ Å. The obtained atomic positions are summarized in Table I. As in the case of zircon ThGeO$_4$, the calculated unit-cell parameters also slightly overestimate the experimental values (see Table I), but the agreement for the atomic positions is quite good. The EOS of scheelite ThGeO$_4$ is given by the following parameters $V_0 = 315.26$ Å$^3$, $B_0 = 173.37$ GPa, and $B_0' = 4.02$. The value of the bulk modulus agrees well with the experimental value (186 GPa) and with our phenomenological estimations (175 GPa). On top of that, in agreement with our experiments, the calculations give a larger compressibility for the $c$-axis than the $a$-axis of scheelite.

Let's concentrate now on the post-scheelite phase of ThGeO$_4$. As pressure increases, our calculations indicate that the scheelite structure becomes unstable against M-fergusonite. This fergusonite structure, a distortion of scheelite [36], only emerges as a structurally different and thermodynamically stable phase above a compression threshold of about 31 GPa. This transition pressure is similar to the experimental value we found for the scheelite-fergusonite transition. At lower pressures, the relaxation of the M-fergusonite structure resulted in the scheelite structure. This is consistent with a quasi-continuous scheelite-to-fergusonite transition with very little volume collapse. This behaviour is also similar to that of most of the ABO$_4$ compounds that undergo the scheelite-fergusonite transition [3]. It should be noted here that, in the range of stability of the fergusonite phase, the energy differences between scheelite and fergusonite are slightly smaller than DFT errors. Indeed, in Fig. 4 it is hard to differentiate both structures. Therefore, to clearly show that the M-fergusonite phase becomes more stable than the scheelite one, we have plotted the energy difference between both structures in the inset of Fig. 4. In spite of this small energy difference, from the total-energy and enthalpy calculations it is found that M-fergusonite becomes the most favourable structure beyond 31 GPa. Therefore, in agreement with our experimental observation, the calculations also suggest that monoclinic fergusonite is the post-scheelite phase of $ThGeO_4$. This conclusion is also consistent with the systematic HP sequence found for orthotungstates, orthomolybdates, and orthovanadates [3, 11]. It is important to note that the high-pressure monoclinic phase reported here has been never found before in germanates. In this regard our results can contribute to a successful anticipation of high-pressure forms in other germanates. In Table II, the calculated structural parameters of the fergusonite phase at 31 GPa are reported. The differences between these parameters and the experimental values are similar to those found in the other two structures of $ThGeO_4$. From our calculations we also determined the EOS of monoclinic $ThGeO_4$. The EOS parameters of this phase are: $V_0 = 315.3$ Å$^3$, $B_0 = 176.16$ GPa, and $B_0' = 4.23$. According with this, the fergusonite and scheelite phase have a very similar compressibility, which is in agreement with the behaviour of other $ABO_4$ compounds [3]. Consequently, the volume change at the phase transition is smaller than 1%, as found in the experiments. Regarding the anisotropy of the M-fergusonite structure, the calculations indicate that the monoclinic distortion increase upon compression. In particular the β angle reaches 90.5º at 40 GPa, and the difference between the unit-cell parameters *a* and *c* increases from 0.02 to 0.05 Å (see Fig. 2). This behaviour is characteristic of the M-fergusonite structure, which distorts upon compression favouring a gradual coordination increase and leading to a pseudo-tetrahedrally coordinated B cation [6].

Based upon first-principle calculations, we have also investigated the evolution of cation–anion distances in ThGeO$_4$. The results obtained for the three phases of interest are summarized in Figure 5. There, it can be seen that in zircon, scheelite, and fergusonite ThGeO$_4$, the Ge-O bonds are much more rigid than the Th-O bonds. This is compatible with the behaviour observed in isostructural compounds [3] and explains why the phenomenological approach developed in Ref. 28 satisfactory estimates the bulk modulus of the different phases of ThGeO$_4$; basically because most of the compression of the crystal comes from the volume reduction of the ThO$_8$ bisdisphenoids. It is interesting also to see that according with our calculations, the distortion of these dodecahedra increases upon compression in the zircon-type phase; the difference between the two Th-O distances is enhanced (see Fig. 5). However, the opposite is true for the scheelite-type phase; i.e. the ThO$_8$ dodechadra become more symmetric. Finally, at the scheelite-fergusonite transition, there is a splitting in the Ge-O distances, resulting in the slightly distorted GeO$_4$ tetrahedra. A similar splitting is also found in the Th-O distances; showing ThO$_8$ units in the M-fergusonite phase existing with four different distances. All these changes are consistent with viewing the scheelite-fergusonite transition as a slight displacement of the atoms, rather than a more dramatic reconstruction of the lattice. Apparently, as observed in ABO$_4$ compounds [3], in ThGeO$_4$ the scheelite-fergusonite transition is caused by small displacements of the Th and Ge atoms from their high-symmetry positions and larger changes in the O positions, which consequently lead to the polyhedral distortion here reported.

## V. CONCLUSIONS

Our X-ray diffraction studies on thorium germanate show that zircon-type ThGeO$_4$ transforms to the scheelite phase around 11 GPa and subsequently to a monoclinic M-fergusonite phase near 26 GPa. This second transition was also detected when compressing samples that crystallize in the scheelite phase at ambient conditions. No additional phase transitions or evidence of decomposition of ThGeO$_4$ were observed up to 40 GPa and on release of pressure ThGeO$_4$ reverts back to scheelite phase without any significant hysteresis. However, the zircon-scheelite transition is non-reversible. The HP M-fergusonite phase is a distorted and compressed version of scheelite obtained by a small distortion of the cation matrix and more significant displacements of the anions. The experimental findings are supported by first-principles calculations performed using the VASP code. From the experiments and the calculations the axial compressibility and the EOS for the different phases of ThGeO$_4$ is also determined, being their compressibility anisotropic. This fact and the determined bulk compressibility can be explained in terms of the different compressibility of the ThO$_8$ and GeO$_4$ polyhedra.

**Acknowledgments:** Financial support from Spanish MALTA-Consolider Ingenio 2010 Program (Project CSD2007-00045) is gratefully acknowledged. This work was partially supported by the Spanish MICCIN under grant MAT2007-65990-C03-01. This work was or portions of this work were performed at HPCAT (Sector 16), Advanced Photon Source (APS), Argonne National Laboratory. HPCAT is supported by DOE-BES, DOE-NNSA, NSF, and the W.M. Keck Foundation. APS is supported by DOE-BES, under Contract No. DE-AC02-06CH11357.

**Table I:** (a) Unit-cell parameters and atomic coordinates of zircon-type ThGeO$_4$ at ambient conditions. The Th atoms are located at the Wyckoff position 4a (0,3/4,1/8), the Ge atoms at 4b (0,1/4,3/8), and the O atoms at 16h (0,$u$,$v$).

|            | $a$ [Å]    | $c$ [Å]    | Atomic coordinates |
|------------|------------|------------|--------------------|
| Experiment | 7.2399(2)  | 6.5416(3)  | $u$ = 0.4308(2)    |
|            |            |            | $v$ = 0.1979(2)    |
| Theory     | 7.3269     | 6.6416     | $u$ = 0.4328       |
|            |            |            | $v$ = 0.1943       |

(b) Unit-cell parameters and atomic coordinates of scheelite-type ThGeO$_4$ at ambient conditions. The Th atoms are located at the Wyckoff position 4a (0,1/4,5/8), the Ge atoms at 4b (0,1/4,1/8), and the O atoms at 16f ($u$,$v$,$w$).

|            | $a$ [Å]    | $c$ [Å]     | Atomic coordinates |
|------------|------------|-------------|--------------------|
| Experiment | 5.1382(6)  | 11.5365(6)  | $u$ = 0.2538(3)    |
|            |            |             | $v$ = 0.1035(3)    |
|            |            |             | $w$ = 0.0454(2)    |
| Theory     | 5.2128     | 11.6022     | $u$ = 0.2565       |
|            |            |             | $v$ = 0.1535       |
|            |            |             | $w$ = 0.0454       |

**Table II:** Structural parameters of fergusonite-type ThGeO$_4$ at 31 GPa. Space group: $I2/a$, Z = 4, a = 4.992 Å, b = 10.982 Å, c = 5.016 Å, β = 90.32°.

|     | Site | x      | y      | z      |
| --- | ---  | ---    | ---    | ---    |
| Th  | 4e   | 0.25   | 0.6253 | 0      |
| Ge  | 4e   | 0.25   | 0.1242 | 0      |
| O$_1$ | 8f   | 0.9136 | 0.9615 | 0.2479 |
| O$_2$ | 8f   | 0.4981 | 0.2117 | 0.8377 |

**Figure Captions**

**Figure 1:** Selection of room-temperature ADXRD data of ThGeO$_4$ at different pressures up to 40 GPa. In all diagrams the background was subtracted. Pressures are indicated in the plot. (r) means pressure release. The ticks indicate the position of the Bragg reflections according with the indexing of the diffraction patterns.

**Figure 2:** Pressure evolution of the unit-cell parameters. Empty symbols: pressure increase. Solid symbols: pressure release. Squares, circles, and triangles represent the zircon, scheelite, and fergusonite phases, respectively. The solid lines are quadratic fits to the experimental data. The dashed lines represent our theoretical results. To facilitate the comparison for the scheelite (fergusonite) phase we plotted *c/2* (*b/2*) instead of *c* (*b*). Note that the crystallographic settings commonly used to describe the scheelite and fergusonite phases are related in such a way that the *c*-axis of the tetragonal unit cell corresponds to the *b*-axis of the monoclinic unit cell

**Figure 3:** Pressure-volume relation in ThGeO$_4$. Empty (solid) Symbols: upstroke (downstroke) experiments. Squares: zircon. Circles: scheelite. Triangle: fergusonite. Solid lines: EOS fit. Dashed (dotted) lines: calculations for the zircon and scheelite (fergusonite) phase.

**Figure 4:** Energy-volume curves calculated for ThGeO$_4$. The structures shown are zircon (solid line), scheelite (dashed line), and fergusonite (dotted line). To better illustrate when fergusonite becomes energetically most stable than scheelite, the inset shows the energy difference between the fergusonite and scheelite phases.

**Figure 5:** Pressure dependence of the interatomic bond distances in the different phases of ThGeO$_4$. For the Ge-O distance, first and second neighbours distances are shown.

**Figure 1**

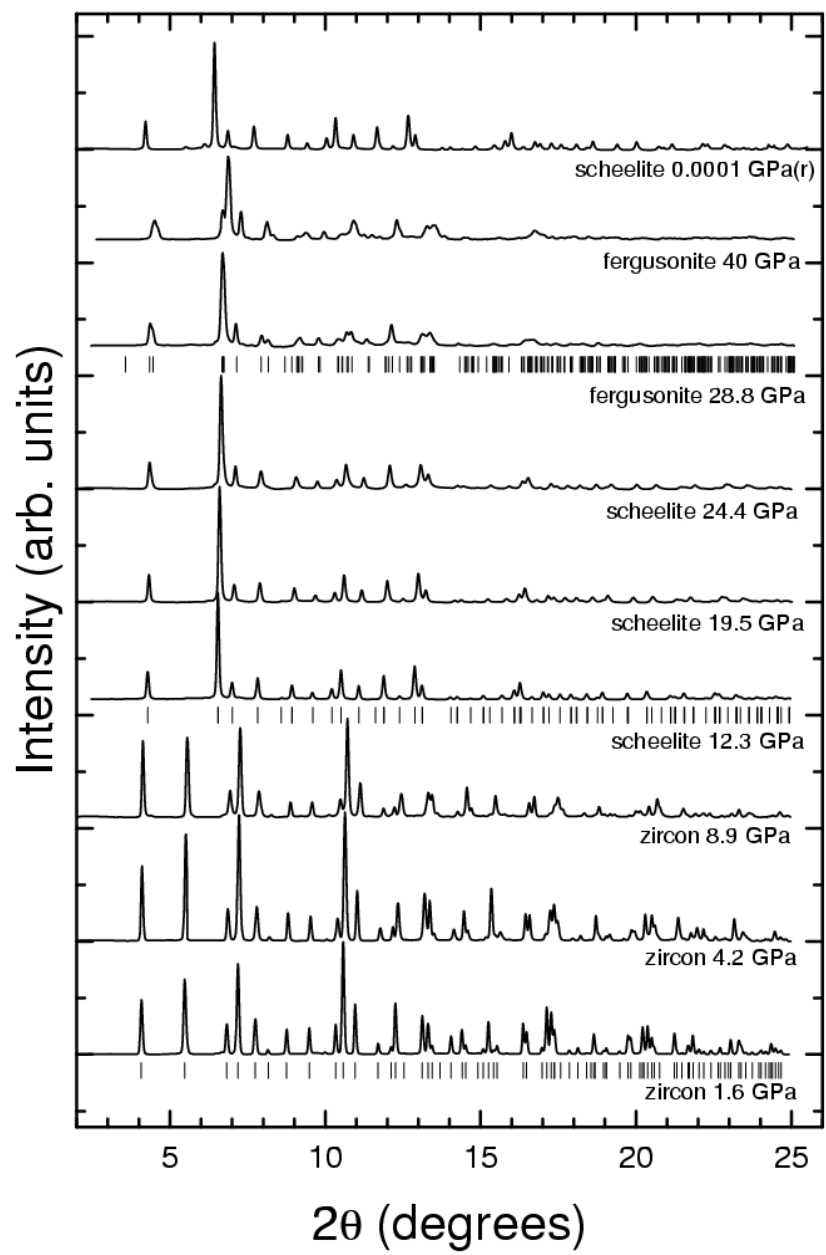

**Figure 2**

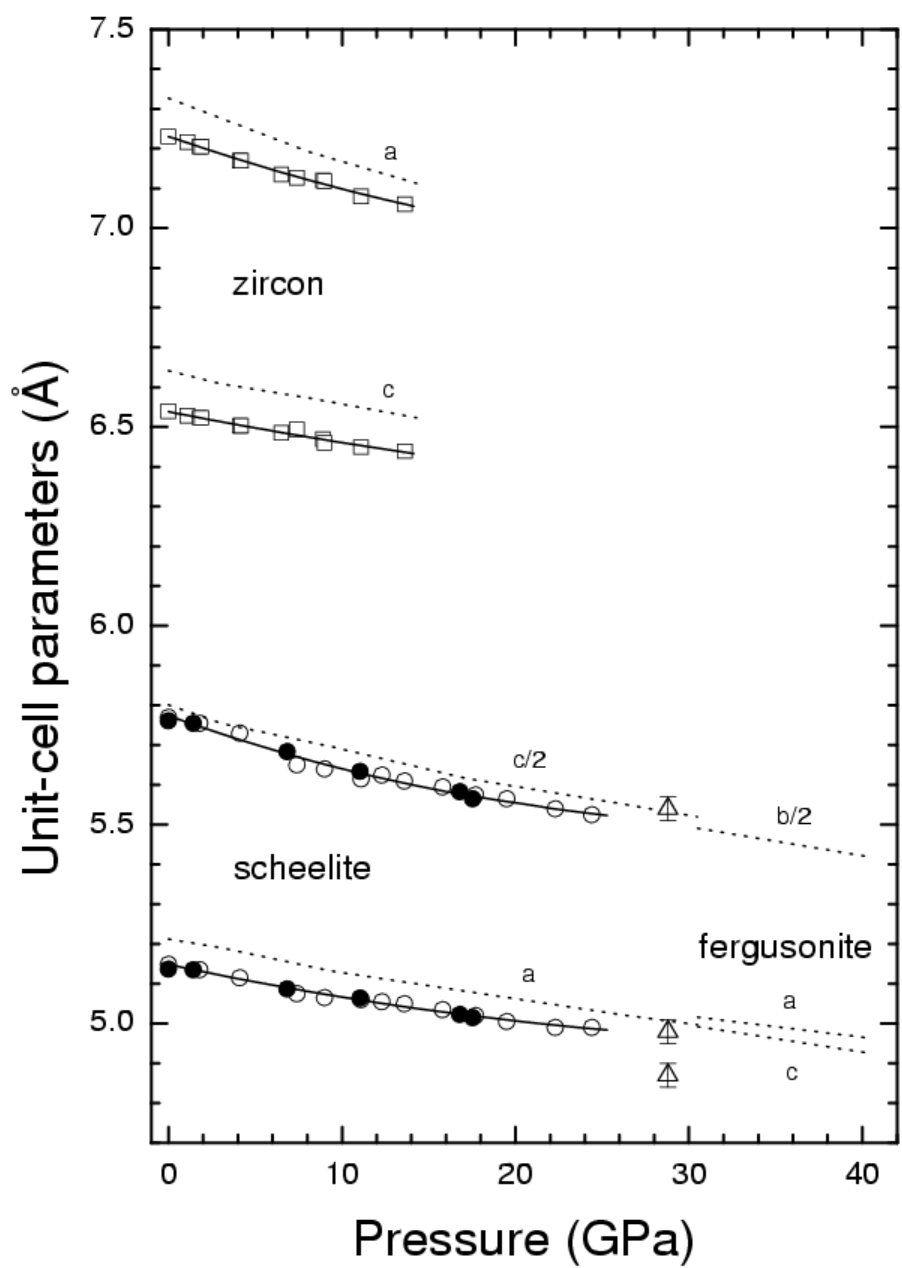

**Figure 3**

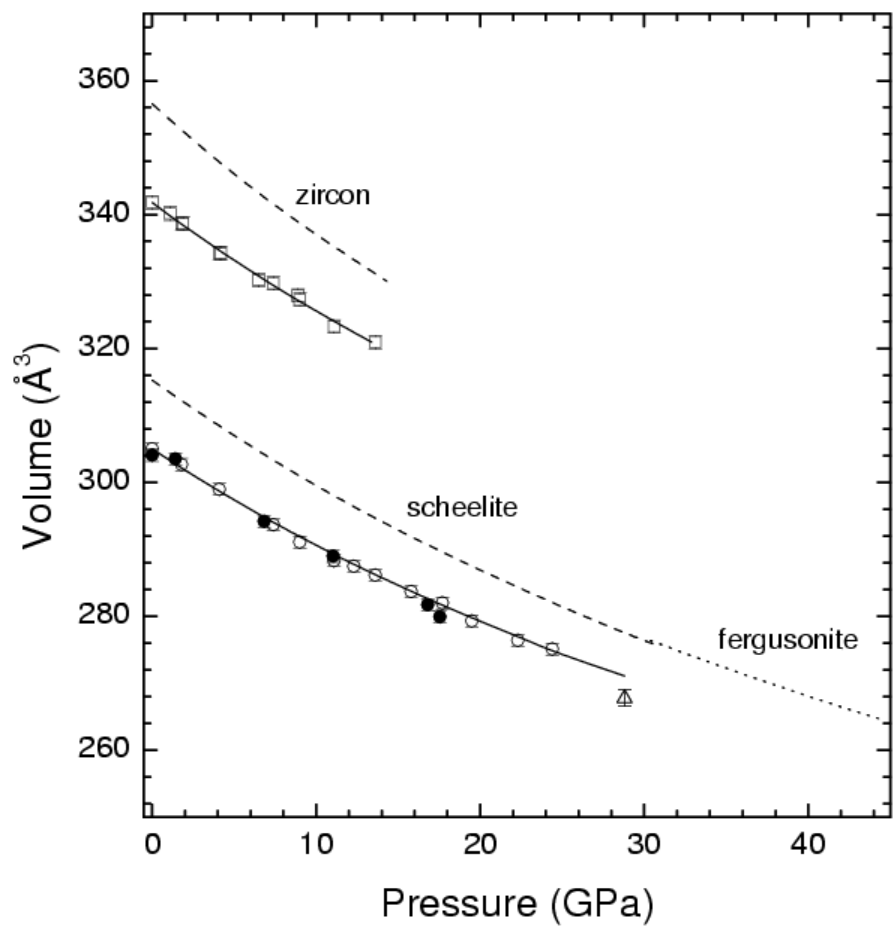

**Figure 4**

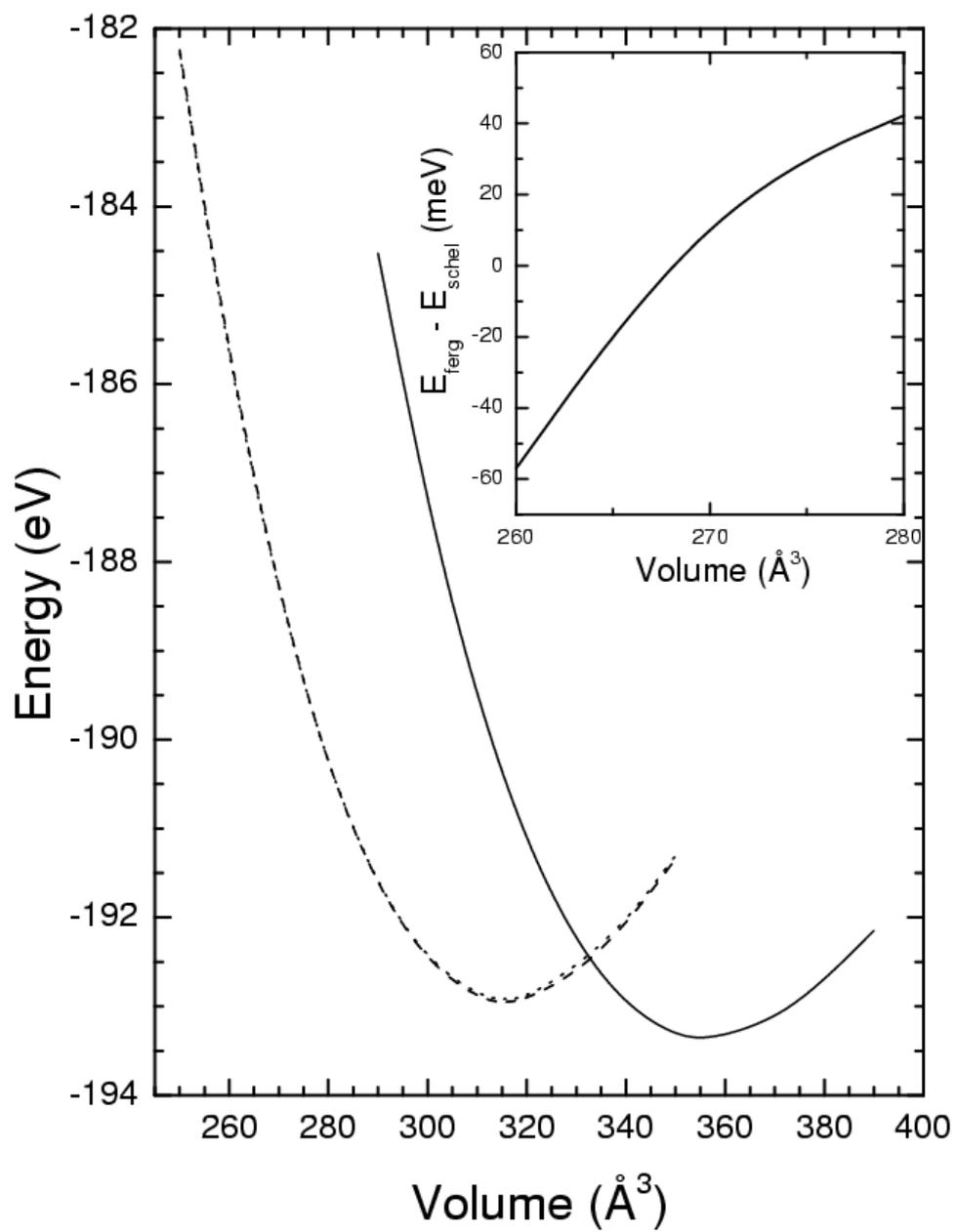

**Figure 5**

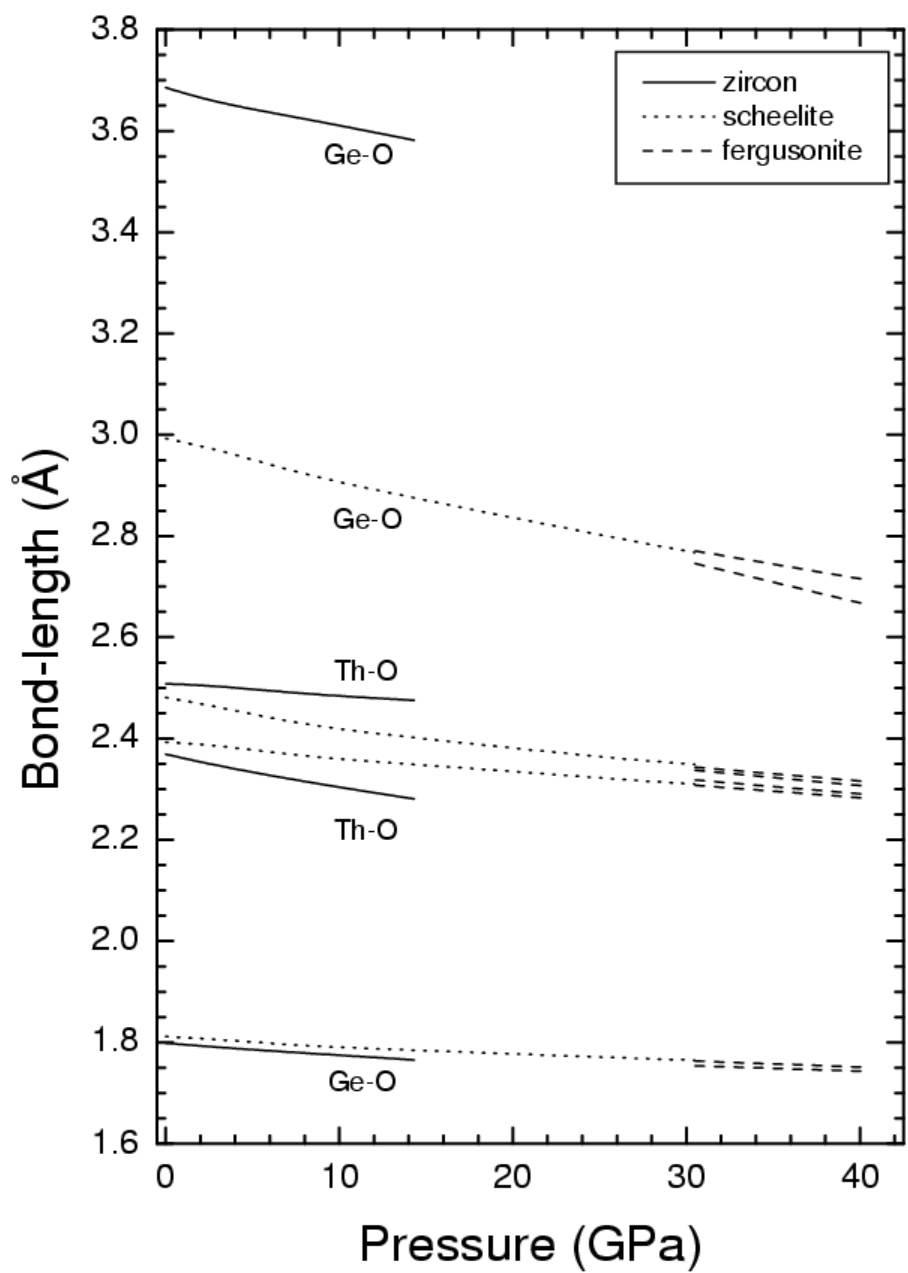